**Title:** BREAST BIOMECANICAL MODELING FOR COMPRESSION OPTIMIZATION IN DIGITAL BREAST TOMOSYNTHESIS

**Authors:** Mîra Anna (1,2), Ann-Katherine Carton (2), Serge Muller (2), Yohan Payan (1)

**Affiliations :** 1. TIMC-IMAG Lab, CNRS & Univ. Grenoble Alpes, France; 2. GE-Healthcare, France;

**Introduction**

Today, projection mammography is the key imaging modality for breast cancer screening and plays an important role in diagnostics. During the exam, the women breast is compressed between two plates until a nearly uniform breast thickness is obtained. Breast compression improves image quality and reduces the absorbed dose of ionizing photons. But breast compression can also be the source of discomfort and sometimes pain for the patient during and after the exam. The discomfort perceived during the exam could deter women from getting the exam. Therefore, an alternative technique with reduced breast compression is of potential interest.

The aim of this work is to develop a biomechanical Finite Element (FE) breast model allowing to investigate alternative breast compression strategies. Ultimately, their impact on mammography image quality and radiation dose could be investigated using a simulation framework generating numerical breast-like phantoms, FE mammographic breast deformation and image acquisitions. The 3D strain/stress cartography derived from the FE solution could be used as a first measure of pain and discomfort.

Previously developed biomechanical breast models have used a simplified breast anatomy by modeling adipose and fibroglandular tissues only [1]. However, breast reconstruction surgery has proven the importance of suspensory ligaments and breast fasciae on breast mechanics [2]. We are considering using a more complex breast anatomy by including breast ligaments as well as the skin and muscles.

A physical correct modeling of the breast deformation requires the knowledge of the stress-free breast configuration; i.e. the absence of gravity load. Here, the stress-free shape was computed using the prediction-correction iterative scheme [3], the unloading procedure used the breast configuration in prone and supine position to find a unique displacement vector field induced by gravitational forces. Finally, the estimated breast deformations under gravity load were compared with the experimental ones obtained from the MR images.

**Materials and Methods**

*Data acquisition and preprocessing.*

The biomechanical breast model is built using MR images of two women between 50 and 55 years old and with respectively small and large breasts. Both women participated on a voluntary basis and signed an informed consent form approved by the ethics committee. The MRI was performed at 3T (Achieva 3.0TX, Philips, NL) at IRMaGe MRI facility (Grenoble, France). Both women were imaged in three configurations: prone, supine and supine tilted positions (Figure 1).

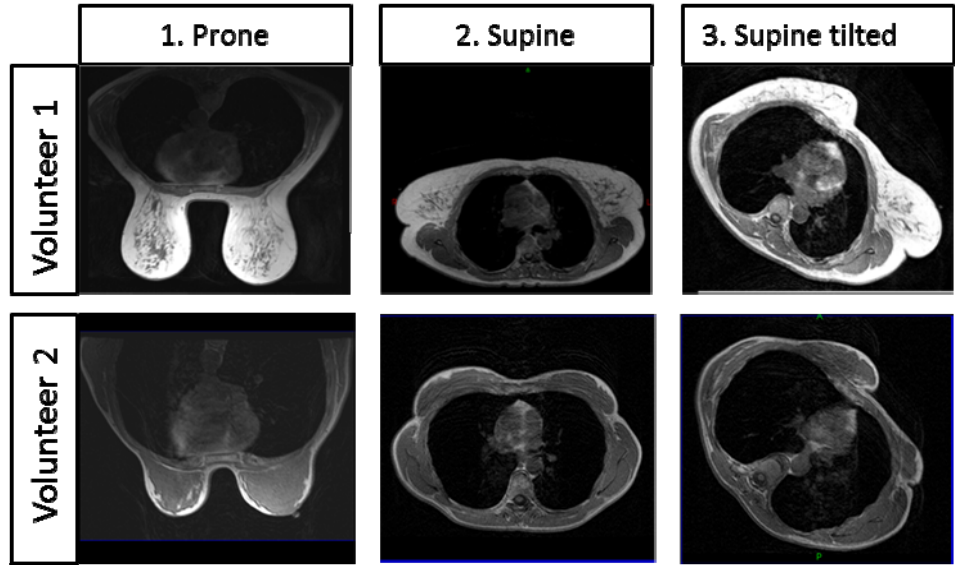

Figure 1: Breast configurations under gravity loads.

The positions were chosen to assess the largest possible breast deformation without any contact area between the patient and the MRI scanner bore. Before image acquisition, ten fiducial markers were fixed on the skin surface; four on the chest wall and six on the breast skin. The markers were used to aid pre-processing registration of the breast volume and to validate the modeled breast deformation.

Image pre-processing starts with an image registration step. The rigid transformation between two breast configurations was computed using the marker's coordinates and the chest wall position in the three images. We assumed that only the four chest-wall markers impact the rigid transformation between breast configurations. Next, the MRI data were classified according to three tissue types; muscle, breast and skin (Figure 2). The segmented data set was then used to reconstruct the 3-dimensional breast geometry. The image preprocessing steps were performed using ITK libraries VERSION_4.3.2 and ITK-Snap software VERSION_3.4.0 [4].

*Patient Specific Finite Element Model (PSFEM)*

From the segmented data, a tetrahedral finite element mesh was generated using ANSYS meshing tool [5]. Two components were created: one representing the breast soft tissues and its surroundings, and the second representing the pectoral muscle (figure 2). The contact surface between the two components was defined as a "no-separation" contact (the two components are bonded) with an infinite friction coefficient [1]. As the skin is a thin layer organ (thickness of 2mm) we used triangular shell elements to represent it.

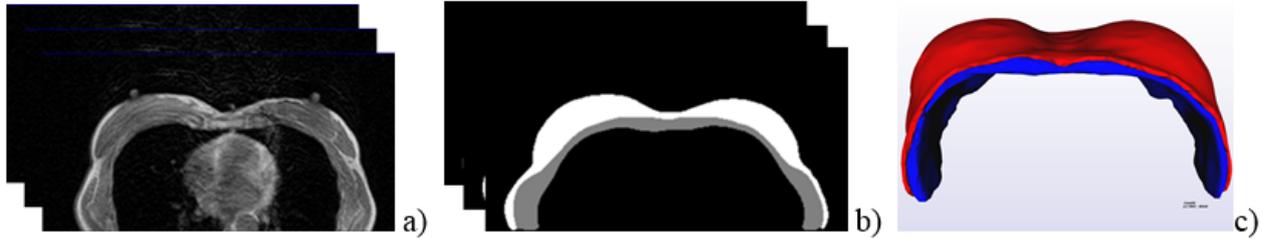

Figure 2: a) MRI data for supine position, b) segmented data, c) corresponding finite element mesh

Finally, the new finite element mesh was subjected to a hyper-elastic quasi-static simulation. Soft tissues were modelled as quasi-incompressible (Poisson Ration = 0.45) neo-Hookean materials. The optimal elastic parameters were determined by minimizing the error between the simulated and the experimental data, using a manual dichotomy on the elastic modulus. The search intervals were defined for each tissue type based on the data found in the previous studies [2,6,7]. The elastic parameter of the pectoral muscle was set to 30kPa and the elastic parameters of breast tissues and skin were varied from 0 to 30kPa and from 5 to 80 kPa respectively.

## Stress-free geometry estimation

In this work, the stress-free geometry was computed using a prediction-correction iterative scheme. The algorithm first proposed by Carter et al. [3] was improved and adapted to our problem.

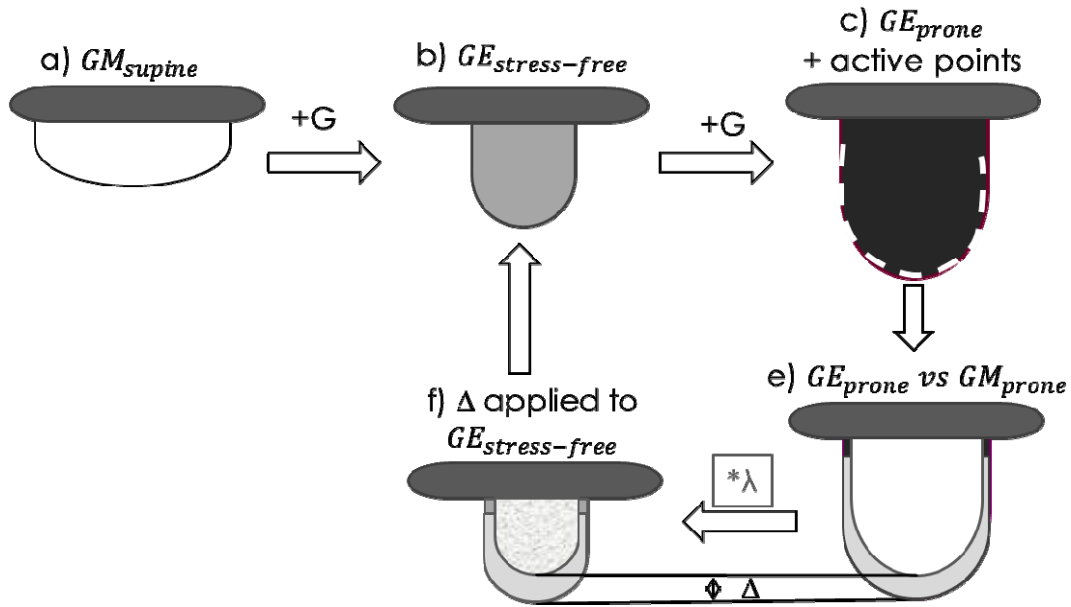

Figure 3: Prediction-correction iterative scheme for stress-free geometry estimation. GM = measured breast geometry; GE= estimated breast geometry. Δ = difference between estimated and measured geometries. λ = regularization factor.

First, we assumed zero internal stress for the breast supine configuration generated from MR images (Figure 3.a Geometry of Measured supine configuration). Then, the gravity load was applied in the reverse direction to give a first estimation of initial stress-free state (Figure 3.b Geometry of Estimated stress-free configuration). Next, the gravity load was applied to the stress-free geometry to estimate the breast geometry in prone configuration (Figure 3. c). At this step, we defined a group of nodes (active nodes) which were used to compute the difference between the nodes position in estimated and measured prone configuration. The nodes were selected at the external surface of the breast. These active node difference was then applied to the stress-free geometry. The process was repeated until convergence was achieved.

**Results**

The breast stress-free geometry was estimated for each pair of elastic parameters defined from the search intervals. After the stress-free geometry estimation, the prone and supine breast positions were computed to evaluate the model. The best fitting of real data in both positions is obtained using an elastic parameter equal to 0.3kPa for breast tissue and equal to 10kPa for skin (see Figure 4). The manual adjustment of elastic parameters conducted to breast tissues far softer than most values found in the literature.

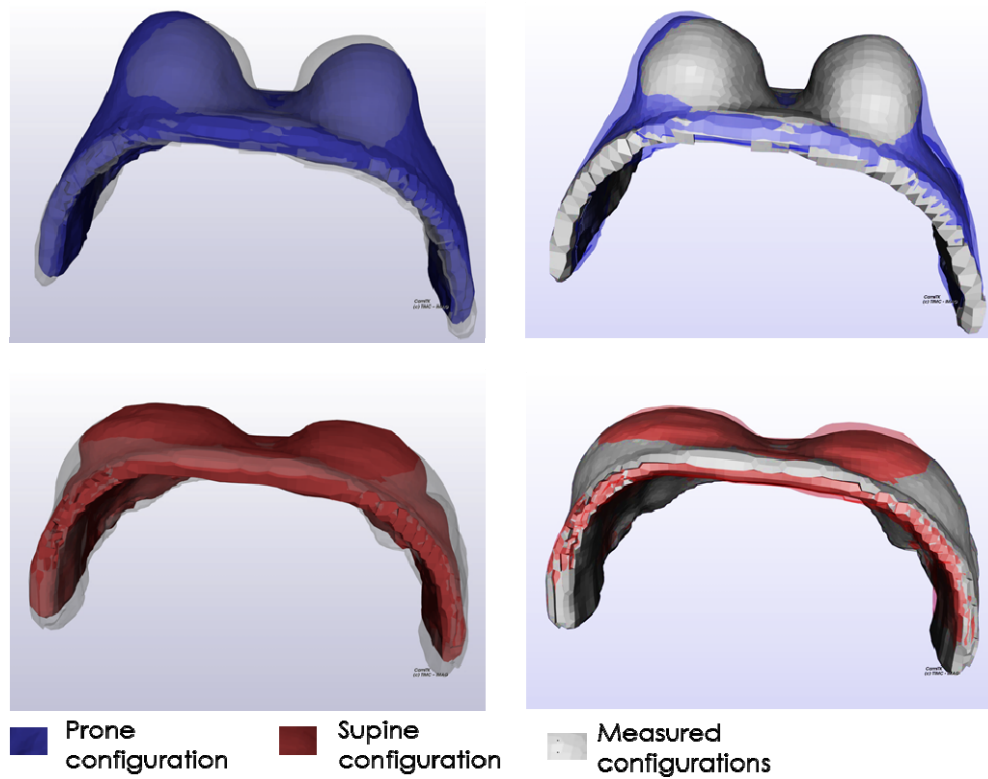

**Figure 4: Comparison of simulated configurations with the measured ones.**

## Discussion

The results derived from the manual adjustment of tissues elasticity revealed that our breast model is highly sensitive on the value of elastic parameters. Therefore, a good estimation of the breast mechanics requires a precise estimation of tissues elasticity for each individual volunteer. We should note however that our results are only consistent with models developed and validated with data using MR images. Most of published papers on breast tissue elasticity propose very high values for fat, glandular and skin elastic parameters [7]. The large difference may come from the inter-women variability of breast elastic properties but also by a wide range of mathematical models and experimental methods.

Knowing that we need a proper estimation of the elastic properties for each volunteer, we will develop an automatic optimization procedure replacing the manual one described here. An automatic optimization should allow a more accurate estimation of elastic parameters. One of the major difficulties of an automatic optimizations is the accumulation of finite elements simulations on the same mesh. As a hyper-elastic model with large deformations is used, the finite element mesh can be distorted after each simulation. Therefore, the optimization process can be stopped due to the poor-quality elements without achieving the convergence.

As a future work, we intend to consider the breast heterogeneity in our model. More specifically, it is necessary to differentiate the glandular tissues and adipose tissues, as well as to introduce the Cooper's ligaments and facias. Our biomechanical model will be then validated using the third breast configuration (supine tilted figure1) and will be used to simulate artificial tomographic images.

## Acknowledgments

This research project is financially supported by ANRT, CIFRE convention n°2014/1357.

We are thanking the IRMaGe MRI facility (Grenoble,France) for their participation in image data acquisition.